\documentstyle[preprint,aps]{revtex}

\begin{document}
\draft
\title{HIGH-MOMENTUM PROTON REMOVAL FROM $^{\bf 16}{\bf O}$ \\
AND THE $\bf (e,e'p)$ CROSS SECTION}
\author{A. Polls}
\address{Departament d'Estructura i Constituents de la Mat\`eria,
Universitat de Barcelona, \\
Diagonal 647, E-8028 Barcelona, Spain}
\author{M. Radici and S. Boffi}
\address{Dipartimento di Fisica Nucleare e Teorica, Universit\`a 
di Pavia, and \\
Istituto Nazionale di Fisica Nucleare, Sezione di Pavia, 
I-27100 Pavia, Italy}
\author{W. H. Dickhoff}
\address{Department of Physics, Washington University, St. Louis,
Missouri 63130}
\author{H. M\"uther}
\address{Institut f\"ur Theoretische Physik, Universit\"at 
T\"ubingen,\\ 
Auf der Morgenstelle 14, D-72076 T\"ubingen, Germany}
\date{\today}
\maketitle
\begin{abstract}
The cross section for the removal of high-momentum protons from
$^{16}{\rm O}$ is calculated for high missing energies.
The admixture of high-momentum nucleons in the $^{16}{\rm O}$
ground state is obtained by calculating the single-hole spectral
function directly in the finite nucleus with the inclusion of 
short-range and tensor correlations induced by a realistic 
meson-exchange interaction. The presence of high-momentum 
nucleons in the transition to final states in $^{15}{\rm N}$ at 
60-100 MeV missing energy is converted to the coincidence cross 
section for the $(e,e'p)$ reaction by including the coupling to 
the electromagnetic probe and the final state interactions of 
the outgoing proton in the same way as in the standard analysis 
of the experimental data. Detectable cross sections for the 
removal of a single proton at these high missing energies are 
obtained which are considerably larger at higher missing 
momentum than the corresponding cross sections for the $p$-wave 
quasihole transitions. Cross sections for these quasihole 
transitions are compared with the most recent experimental data 
available.
\end{abstract}

\narrowtext

\section{Introduction}
Experimental progress in the exclusive $(e,e'p)$ reaction in recent
years has provided a clear picture of the limitations of the simple
shell-model description of closed-shell nuclei.
Of particular interest is the reduction of the single-particle (sp)
strength for the removal of particles with valence hole
quantum numbers with respect to the simple shell-model estimate
which corresponds to a spectroscopic factor of 1 for such states.
Typical experimental results\cite{data} for closed-shell nuclei 
exhibit reductions of about 30\% to 45\% for these spectroscopic
factors. In the case of ${}^{208}{\rm Pb}$, one obtains a 
spectroscopic factor for the transition to the ground state of 
${}^{207}{\rm Tl}$ of about 0.65 which is associated with the 
removal of a $3s{1 \over 2}$ proton. An analysis which uses 
information obtained from elastic electron scattering, indicates 
that the total occupation number for this state is about 10\%
higher\cite{wagner}, corresponding to 0.75. This additional 
background strength should be present at higher missing energy
and is presumed to be highly fragmented. The depletion of more 
deeply bound orbitals is expected to be somewhat less as
suggested by theoretical considerations\cite{dimu92} which also 
indicate that the strength in the background, outside the main 
peak, corresponds to about 10\% (see also \cite{mah91}).

Recent experimental results for ${}^{16}{\rm O}$\cite{leus94} 
yield a combined quasihole strength for the $p{1 \over 2}$ and 
$p {3 \over 2}$ states corresponding to about 65\% with the 
$p{1 \over 2}$ strength concentrated in one peak and the 
$p{3 \over 2}$ strength fragmented already over several peaks. 
Recent theoretical results yield about 76\% for these $p$ 
states\cite{geu96} without reproducing the fragmentation of the 
$p{3 \over 2}$ strength. This calculation includes the influence 
of both long-range correlations, associated with a large 
shell-model space, as well as short-range correlations.
Although the inclusion of long-range correlations yields a good
representation of the $l=2$ strength, it fails to account for the
presence of positive parity fragments below the first $p{3\over 2}$
fragment. This suggests that additional improvement of the 
treatment of long-range correlations is indicated possibly 
including a correct treatment of the center-of-mass 
motion\cite{rad94}. The contribution to the depletion of the sp 
strength due to short-range correlations is typically about 10\%.
This result is obtained both in nuclear matter calculations, as 
reviewed in \cite{dimu92}, and in calculations directly for finite
(medium-)heavy nuclei\cite{rad94,mudi94,mu95,po95,geu96}. 
Although the influence of long-range correlations on the 
distribution of the sp strength is substantial, it is clear that 
a sizable fraction of the missing sp strength is due to 
short-range effects. The experimental data\cite{data,leus94} 
indicate that only about 70\% of the expected protons in the 
nucleus has been detected in the energy and momentum domain 
studied so far. It is therefore important to establish precisely 
where the protons which have been admixed into the nuclear 
ground state due to short-range and tensor correlations, can be 
detected in the $(e,e'p)$ reaction and with what cross section. 

The influence of short-range correlations on the presence of 
high-momentum components in finite (medium-)heavy nuclei has been 
calculated in \cite{mudi94,mu95,po95}. In this work the spectral 
function for ${}^{16}{\rm O}$ has been calculated from a 
realistic interaction without recourse to some form of local 
density approximation\cite{sic94,vnec95}. No substantial 
high-momentum components are obtained in \cite{mudi94,mu95,po95} 
at small missing energy. With increasing missing energy, however, 
one recovers the high-momentum components which have been admixed 
into the ground state. The physics of these features can be 
traced back to the realization that the admixture of 
high-momenta requires the coupling to two-hole-one-particle 
(2h1p) states in the self-energy for a nucleon with high 
momentum. In nuclear matter the conservation of momentum requires 
the equality of the 2h1p momentum in the self-energy and the
external high momentum. Since the two-hole state has a relatively 
small total pair momentum, one automatically needs an essentially 
equally large and opposite momentum for the intermediate 
one-particle state to fulfill momentum conservation. As a result, 
the relevant intermediate 2h1p states will lie 
at increasing excitation energy with increasing momentum. 
Considerations of this type are well known for nuclear matter 
(see {\it e.g.} \cite{cio91}), but are approximately valid in 
finite nuclei as well. Recent experiments on 
${}^{208}{\rm Pb}$\cite{bob94} and ${}^{16}{\rm O}$\cite{blo95} 
essentially confirm that the presence of high-momentum components 
in the quasihole states accounts for only a tiny fraction of
the sp strength.

The theoretical prediction concerning the presence of high-momentum
components at high missing energy remains to be verified 
experimentally, however. In order to facilitate and support 
these efforts, the present work aims to combine the calculation 
of the spectral function at these energies with the description  
of both the electromagnetic vertex and final state 
interactions (FSI) in order to produce realistic estimates of the 
exclusive $(e,e'p)$ cross section under experimental conditions 
possible at NIKHEF and Mainz. The impulse approximation has been 
adopted for the electromagnetic current operator, which describes 
the nonrelativistic reduction (up to fourth order in the inverse 
nucleon mass~\cite{gp80}) of the coupling between the external 
virtual photon and single nucleons only. The treatment of FSI has 
been developed by the Pavia 
group~\cite{bc81,bccgp82,bgp82,br91,bgprep} (see also 
Ref.~\cite{libro}) and takes into account the average complex 
optical potential the nucleon experiences on its way out of the 
nucleus. Other contributions to the exclusive $(e,e'p)$ reaction 
are present in principle, such as two-step mechanisms in the 
final state or the decay of initial collective excitations in the 
target nucleus. However, by transferring sufficiently high energy 
$\omega$ to the target nucleus and by selecting typical 
kinematical conditions corresponding to the socalled quasielastic 
peak with $\omega = q^2/2m$ ($q$ the momentum transfer and 
$m$ the nucleon mass), these contributions are 
suppressed. In these conditions, adopted in the most recent 
experiments, the direct knockout mechanism has been shown to be 
the dominant contribution~\cite{bgprep} and essentially 
corresponds to calculating the combined probability for exciting a
correlated particle (which is ultimately detected) and a 
correlated hole such that energy and momentum are conserved but 
no further interaction of the particle with the hole is included.

The calculation of the spectral function for ${}^{16}{\rm O}$ is
reviewed in Sec. II. Special attention is given to a separable
representation of the spectral function which facilitates the 
practical implementation of the inclusion of FSI. In Sec. III the 
general formalism of the Distorted Wave Impulse Approximation 
(DWIA) is briefly reviewed. The influence of the FSI is studied 
in Sec. IV for the quasihole transitions for which data are 
available\cite{leus94,blo95}. Extending the calculation of the 
cross section to higher missing energies yields the expected 
rise of high missing-momentum components in the cross section 
in comparison to the results near the Fermi energy. The 
contribution of various partial waves is studied demonstrating the
increasing importance of higher $l$-values with increasing missing
momentum. All these results are discussed in Sec. IV and a brief 
summary is presented in Sec. V.

\section{the single-particle spectral function}
The calculation of the cross section for exclusive $(e,e'p)$ 
processes requires the knowledge of the hole spectral function 
which is defined in the following way
\begin{eqnarray}
S({\bf p},m_s,m_{\tau},{\bf p'},m_s^{'},m_{\tau};E)&= & \sum_n 
\left \langle \Psi_0^{\rm A} \mid 
a^{\dagger}({\bf p'},m_s',m_{\tau}) 
\mid \Psi_n^{{\rm A} - 1}\right \rangle
\left \langle \Psi_n^{{\rm A} - 1} \mid 
a({\bf p},m_s,m_{\tau}) \mid \Psi_0^{\rm A} \right \rangle 
\nonumber \\
 & & \delta(E-(E_0^{\rm A}-E_n^{{\rm A} - 1})), \label{eq:spec}
\end{eqnarray}
where the summation over $n$ runs over the discrete excited 
states as well as over the continuum of the (A-1) particle 
system, $\left |\Psi_0^{\rm A} \right\rangle $ is the ground 
state of the initial nucleus and $a({\bf p},m_s,m_{\tau})$ 
$(a^{\dagger}({\bf p'},m_s',m_{\tau}))$ is the annihilation 
(creation) operator with the specified sp quantum numbers for 
momenta and third component of spin and isospin, respectively. 
The spectral function is diagonal in the third component of the 
isospin and ignoring the Coulomb interaction between the protons,
the spectral functions for protons and neutrons are identical 
for N=Z nuclei. Therefore in the following we have dropped the 
isospin quantum number $m_{\tau}$. Note that the energy variable 
$E$ in this definition of the spectral function refers to minus 
the excitation energy of state $n$ in the A-1 particle system 
with respect to the ground-state energy $(E_0^{\rm A})$ of the 
nucleus with A nucleons.

To proceed further in the calculations it is useful to introduce a
partial wave decomposition which yields the spectral function 
for a nucleon in the sp basis with orbital angular momentum $l$,
total angular momentum $j$, and momentum $p$
\begin{equation}
S_{lj}(p,p';E)= \sum_n \left \langle \Psi_0^{\rm A} \mid 
a^{\dagger}_{p'lj} \mid \Psi_n^{{\rm A} - 1}\right \rangle  
\left \langle  \Psi_n^{{\rm A} - 1} \mid a_{plj} \mid 
\Psi_0^{\rm A} \right \rangle  
\delta(E-(E_0^{\rm A}-E_n^{{\rm A} - 1})), \label{eq:specl}
\end{equation}
where $a_{plj}$($a^{\dagger}_{p'lj}$) denotes the corresponding
removal (addition) operator. The spectral functions for the 
various partial waves, $S_{lj}(p,p';E)$, have been obtained from 
the imaginary part of the corresponding sp propagator 
$g_{lj}(p,p';E)$. This Green's function solves the Dyson equation
\begin{equation}
g_{lj}(p_1,p_2;E)=g_{lj}^{(0)}(p_1,p_2;E)+\int dp_3 \int dp_4
g_{lj}^{(0)}(p_1,p_3;E) \Delta \Sigma_{lj}(p_3,p_4;E) 
g_{lj}(p_4,p_2;E), \label{eq:dyson}
\end{equation}
where $g^{(0)}$ refers to a Hartree-Fock propagator and
$\Delta\Sigma_{lj}$ represents contributions to the real and 
imaginary parts of the irreducible self-energy, which go beyond 
the Hartree-Fock approximation of the nucleon self-energy used to 
derive $g^{(0)}$. Although the evaluation of the self-energy as 
well as the solution of the Dyson equation has been discussed in 
detail in previous publications \cite{mu95,po95} we include 
here a brief summary of the relevant aspects of the method.

\subsection  {Calculation of the nucleon self-energy}
The self-energy is evaluated in terms of a $G$-matrix which is 
obtained as a solution of the Bethe-Goldstone equation for 
nuclear matter choosing for the bare NN interaction the 
one-boson-exchange potential B defined by Machleidt 
(Ref. \cite{rupr}, Table A.2). The Bethe-Goldstone equation has 
been solved for a Fermi momentum $k_{\rm F} = 1.4 \  
{\rm fm}^{-1}$ and starting energy $-10$ MeV. The choices for the 
density of nuclear matter and the starting energy are rather 
arbitrary. It turns out, however, that the calculation of the 
Hartree-Fock term (Fig. 1a) is not very sensitive to this 
choice \cite{bm2}. Furthermore, we will correct this nuclear 
matter approximation by calculating the 
two-particle-one-hole (2p1h) term displayed in Fig. 1b directly 
for the finite system. This second-order correction, which 
assumes harmonic oscillator states for the occupied (hole) 
states and plane waves for the intermediate unbound particle 
states, incorporates the correct energy and density dependence 
characteristic of a finite nucleus $G$-matrix. To evaluate the 
diagrams in Fig. 1, we need matrix elements in a mixed 
representation of one particle in a bound harmonic oscillator 
while the other is in a plane wave state. Using vector bracket 
transformation coefficients \cite{vecbr} one can transform matrix 
elements from the representation in coordinates of relative and 
center-of-mass momenta to the coordinates of sp momenta in the 
laboratory frame in which the two particle state is described by
\begin{equation}
\left | p_1 l_1 j_1 p_2 l_2j_2 J T \right \rangle  
\label{eq:twostate}
\end{equation}
where $p_i$, $l_i$ and $j_i$ refer to momentum and angular 
momenta of particle $i$ whereas $J$ and $T$ define the total 
angular momentum and isospin of the two-particle state. 
Performing an integration over one of the $p_i$, one obtains a 
two-particle state in the  mixed representation,  
\begin{equation}
\left | n_1 l_1 j_1 p_2 l_2 j_2 J T \right \rangle = 
\int_0^{\infty} dp_1 p_1^2 R_{n_1,l_1}(\alpha p_1) \left | 
p_1 l_1 j_1 p_2 l_2 j_2 J T \right \rangle .
\label{eq:labstate}
\end{equation}
Here $R_{n_1,l_1} $ stands for the radial oscillator function 
and the oscillator length $\alpha = $ 1.72 fm$^{-1}$ has been 
chosen to have an appropriate description of the bound sp states 
in $^{16}$O. Using the notation defined in 
Eqs.~(\ref{eq:twostate}) and (\ref{eq:labstate}), our 
Hartree-Fock approximation for the self-energy is obtained in the 
momentum representation,
\begin{equation}
\Sigma_{l_1j_1}^{\rm HF}(p_1, p_1') = {{1}\over {2(2j_1+1)}}
  \sum_{n_2 l_2 j_2 J T}   (2J+1)(2T+1)  
\left \langle p_1 l_1 j_1 n_2 l_2 j_2 J T \mid G \mid 
p_1' l_1 j_1 n_2 l_2 j_2 J T \right \rangle .
\label {eq:hf}
\end{equation}
The summation over the oscillator quantum numbers is restricted 
to the states occupied in the independent particle model of 
$^{16}$O. This Hartree-Fock part of the self-energy is real and 
does not depend on the energy.

The terms of lowest order in $G$ which give rise to an imaginary 
part in the self-energy are represented by the diagrams 
displayed in Figs.\ \ref{fig:diag}b and \ref{fig:diag}c, 
referring to intermediate 2p1h and 2h1p states respectively. 
The 2p1h contribution to the imaginary part is given by
\begin{eqnarray}
{W}^{\rm 2p1h}_{l_1j_1} (p_1,p'_1; E) =& 
{\displaystyle {-1 \over {2(2j_1+1)}}} \> \>
{ {\lower7pt\hbox{$_{n_2 l_2 j_2}$}} \kern-20pt 
{\hbox{\raise2.5pt \hbox{$\sum$}}} } \quad 
{ {\lower7pt\hbox{$_{l L}$}} \kern-10pt 
{\hbox{\raise2.5pt \hbox{$\sum$}}} } \>
{ {\lower7pt\hbox{$_{J J_S S T}$}} \kern-22pt 
{\hbox{\raise2.5pt \hbox{$\sum$}}} } \quad
\int k^2 dk \int K^2 dK  (2J+1) (2T+1) \nonumber \\
& \times \left\langle p_1 l_1 j_1 n_2 l_2 j_2 J T 
\right| G \left| k l S J_S K L T \right\rangle
\left\langle k l S J_S K L T 
\right| G \left| p'_1 l_1 j_1 n_2 l_2 j_2 J T \right\rangle 
\nonumber \\
& \times \pi \delta\left(E + \epsilon_{n_2 l_2 j_2} - 
{\displaystyle {K^2 \over 4m}} - {\displaystyle {k^2 \over m}} \right), 
\label{eq:w2p1h}
\end{eqnarray}
where the ``experimental'' sp energies $\epsilon_{n_2 l_2 j_2}$
are used for the hole states ($-47$ MeV, $-21.8$ MeV, $-15.7$ MeV for 
$s {1 \over 2}$, $p {3 \over 2}$ and $p {1 \over 2}$ states, 
respectively), while the energies of the particle states are 
given in terms of the kinetic energy only. The plane waves
associated with the particle states in the intermediate states 
are properly orthogonalized to the bound sp states following the 
techniques discussed by Borromeo  et al. \cite{boro}. The 2h1p 
contribution to the imaginary part 
${W}^{\rm 2h1p}_{l_1j_1}(p_1,p'_1; E)$ can be calculated in a 
similar way (see also \cite{boro}).

Our choice to assume pure kinetic energies for the particle 
states in calculating the imaginary parts of $W^{\rm 2p1h}$ 
(Eq.~(\ref{eq:w2p1h})) and $W^{\rm 2h1p}$ may not be very 
realistic for the excitation modes at low energy. Indeed a 
sizable imaginary part in $W^{\rm 2h1p}$ is obtained only for 
energies $E$ below $-40$ MeV. As we are mainly interested, however, 
in the effects of short-range correlations, which lead to 
excitations of particle states with high momentum, the choice 
seems to be appropriate. A different approach would be required 
to treat the coupling to the very low-lying 2p1h and 2h1p states 
in an adequate way. Attempts at such a treatment can be found in
Refs. \cite{brand,rijsd,skou1,skou2,nili96,geu96}. The 2p1h 
contribution to the real part of the self-energy can be calculated
from the imaginary part $W^{\rm 2p1h}$ using a dispersion relation
\cite{mbbd}
\begin{equation}
V^{\rm 2p1h}_{l_1j_1}(p_1,p_1';E) = {1 \over \pi} \  
{\cal P} \int_{-\infty}^{\infty} 
{{W^{\rm 2p1h}_{l_1j_1}(p_1,p_1';E')} \over {E'-E}} dE', 
\label{eq:disper1}
\end{equation}
where ${\cal P}$ represents a principal value integral. A similar 
dispersion relation holds for $V^{\rm 2h1p}$ and $W^{\rm 2h1p}$.
 
Since the Hartree--Fock contribution $\Sigma^{\rm HF}$ has been 
calculated in terms of a nuclear matter $G$-matrix, it already 
contains 2p1h terms of the kind displayed in Fig.\ 
\ref{fig:diag}b. In order to avoid such an overcounting of the 
particle-particle ladder terms, we subtract from the real part of 
the self-energy a correction term ($V_{\rm c}$), which just 
contains the 2p1h contribution calculated in nuclear matter. 
Summing up the various contributions we obtain for the 
self-energy the following expressions
\begin{equation}
\Sigma  = \Sigma^{\rm HF} + \Delta\Sigma = \Sigma^{\rm HF} + 
\left( V^{\rm 2p1h} - V_{\rm c} + V^{\rm 2h1p} \right)
+ \left( W^{\rm 2p1h} + W^{\rm 2h1p} \right) . \label{eq:defsel}
\end{equation}

\subsection  {Solution of the Dyson equation }
The next step is to solve the Dyson equation (\ref{eq:dyson}) for 
the sp propagator. To this aim, we discretize the integrals in 
this equation by  considering a complete basis within a spherical 
box of a radius $R_{\rm box}$. The calculated observables are 
independent of the choice of $R_{\rm box}$, if it is chosen to be 
around 15 fm or larger. A complete and orthonormal set of regular 
basis functions within this box is given by
\begin{equation}
\Phi_{iljm} ({\bf r}) = \left\langle {\bf r} \vert p_i l j m
\right\rangle = N_{il} \  j_l(p_i r) \  {\cal Y}_{ljm} 
(\theta, \phi) . \label{eq:boxbas}
\end{equation}
In this equation ${\cal Y}_{ljm}$ represent the spherical 
harmonics including the spin degrees of freedom and $j_l$ denote 
the spherical Bessel functions for the discrete momenta $p_i$ 
which fulfill
\begin{equation}
j_l (p_i R_{\rm box}) = 0 .
\label{eq:bound}
\end{equation}

Note that the basis functions defined for discrete values of the 
momentum $p_i$ within the box differ from the plane wave states 
defined in the continuum with the corresponding momentum just by 
the normalization constant, which is 
$\sqrt{\textstyle{2 \over \pi}}$ for the latter. This enables us 
to determine the matrix elements of the nucleon self-energy in 
the basis of Eq.~(\ref{eq:boxbas}) from the results presented in 
the preceding Subsection.

As a first step we determine the Hartree-Fock approximation for the
sp Green's function in the ``box basis''. For that purpose the
Hartree-Fock Hamiltonian is diagonalized
\begin{equation}
\sum_{n=1}^{N_{\rm max}} \left\langle p_i \right| 
\frac{p_i^2}{2m} \delta_{in} + \Sigma^{\rm HF}_{lj} \left| p_n 
\right\rangle \left\langle p_n \vert \alpha \right\rangle_{lj} = 
\epsilon^{\rm HF}_{\alpha lj} \left\langle p_i \vert 
\alpha \right\rangle_{lj}. \label{eq:hfequ}
\end{equation}
Here and in the following the set of basis states in the box has 
been truncated by assuming an appropriate $N_{\rm max}$. In the 
basis of Hartree-Fock states $\left| \alpha \right\rangle$, the 
Hartree-Fock propagator is diagonal and given by
\begin{equation}
g_{lj}^{(0)} (\alpha; E) = 
\frac{1}{E-\epsilon^{\rm HF}_{\alpha lj} \pm i\eta} , 
\label{eq:green0}
\end{equation}
where the sign in front of the infinitesimal imaginary quantity 
$i\eta$ is positive (negative) if $\epsilon^{\rm HF}_{\alpha lj}$ 
is above (below) the Fermi energy. With these ingredients one can 
solve the Dyson equation (\ref{eq:dyson}). One possibility is to 
determine first the socalled reducible self-energy, originating 
from an iteration of $\Delta\Sigma$, by solving
\begin{equation}
\left\langle \alpha \right| \Sigma^{\rm red}_{lj}(E) \left| 
\beta \right\rangle =
\left\langle \alpha \right| \Delta\Sigma_{lj}(E) \left| \beta 
\right\rangle
+ \sum_\gamma
\left\langle \alpha \right| \Delta\Sigma_{lj}(E) \left| \gamma 
\right\rangle
g_{lj}^{(0)} (\gamma ; E) \left\langle \gamma \right| 
\Sigma^{\rm red}_{lj}(E) \left| \beta \right\rangle
\end{equation}
and obtain the propagator from
\begin{equation}
g_{lj} (\alpha ,\beta ;E ) = \delta_{\alpha ,\beta} \  
g_{lj}^{(0)} (\alpha ;E ) + g_{lj}^{(0)} (\alpha ;E )
\left\langle \alpha \right| \Sigma^{\rm red}_{lj}(E) \left| 
\beta \right\rangle g_{lj}^{(0)} (\beta ;E ) .
\end{equation}
Using this representation of the Green's function one can 
calculate the spectral function in the ``box basis'' from
\begin{equation}
\tilde S_{lj}^{\rm c} (p_m,p_n; E) = \frac{1}{\pi} \  \mbox{Im} 
\left( \sum_{\alpha, \beta} \left\langle p_m \vert \alpha 
\right\rangle_{lj} \  g_{lj} (\alpha ,\beta ;E )
\left\langle \beta \vert p_n \right\rangle_{lj}\right) . 
\label{eq:skob}
\end{equation}
For energies $E$ below the lowest sp energy of a
given Hartree-Fock state (with $lj$)
this spectral function is different from zero
only due to the imaginary part in $\Sigma^{\rm red}$.
This contribution involves the coupling to the continuum of 2h1p 
states and is therefore nonvanishing only for energies at which 
the corresponding irreducible self-energy $\Delta\Sigma$ has a 
non-zero imaginary part. Besides this continuum contribution, the 
hole spectral function also receives contributions from the 
quasihole states \cite{mu95}. The energies and wave functions of 
these quasihole states can be determined by diagonalizing the 
Hartree-Fock Hamiltonian plus $\Delta\Sigma$ in the ``box basis''
\begin{equation}
\sum_{n=1}^{N_{\rm max}} \left\langle p_i \right| 
\frac{p_i^2}{2m} \delta_{in} + \Sigma^{\rm HF}_{lj} + 
\Delta\Sigma_{lj} (E=\epsilon^{\rm qh}_{\Upsilon lj})
\left| p_n \right\rangle \left\langle p_n \vert \Upsilon 
\right\rangle_{lj} = \epsilon^{\rm qh}_{\Upsilon lj} \  
\left\langle p_i \vert \Upsilon \right\rangle_{lj}. 
\label{eq:qhequ}
\end{equation}
Since in the present work $\Delta\Sigma$ only contains a sizable 
imaginary part for energies $E$ below 
$\epsilon^{\rm qh}_{\Upsilon}$, the energies of the quasihole 
states are real and the continuum contribution to the spectral 
function is separated in energy from the quasihole contribution. 
The quasihole contribution to the hole spectral function is
given by
\begin{equation}
\tilde S^{\rm qh}_{\Upsilon lj} (p_m,p_n; E) = Z_{\Upsilon lj} 
{\left\langle p_m \vert \Upsilon \right\rangle_{lj} 
\left\langle\Upsilon \vert p_n \right\rangle_{lj}}
 \, \delta (E - \epsilon^{\rm qh}_{\Upsilon lj}), \label{eq:skoqh}
\end{equation}
with the spectroscopic factor for the quasihole state given by 
\cite{mu95}
\begin{equation}
Z_{\Upsilon lj} =
\bigg( {1-{\partial \left\langle \Upsilon \right| 
\Delta\Sigma_{lj}(E) \left| \Upsilon \right\rangle \over
\partial E} \bigg|_{\epsilon^{\rm qh}_{\Upsilon lj}}} \bigg)^{-1} .
\label{eq:qhs}
\end{equation}
Finally, the continuum contribution of Eq.~(\ref{eq:skob}) and 
the quasihole parts of Eq.~(\ref{eq:skoqh}), which are obtained 
in the basis of box states, can be added and renormalized to 
obtain the spectral function in the continuum representation at 
the momenta defined by Eq.~(\ref{eq:bound})
\begin{equation}
S_{lj} (p_m,p_n;E) = \frac{2}{\pi} \  \frac{1}{N_{il}^2} \bigl( 
\tilde S^{\rm c}_{lj} (p_m,p_n;E) + \sum_{\Upsilon} \tilde 
S^{\rm qh}_{\Upsilon lj} (p_m,p_n;E) \bigr).
\label{eq:renor}
\end{equation}
It is useful to have a separable representation of the spectral 
function in momentum space. For a given energy, the spectral 
function in the box is represented by a matrix in momentum space; 
after diagonalizing this matrix one obtains
\begin{equation}
S_{lj}(p_m,p_n;E)= \sum_{i}^{N_{max}} S_{lj} (i) \  \phi_i(p_m) 
\  \phi_i(p_n)
\label{eq:separ}
\end{equation}
where $S_{lj}(i)$ are the eigenvalues and $\phi_i$ are the 
corresponding eigenfunctions. In all cases considered here, it is 
enough to consider the first 5 or 6 largest eigenvalues in 
Eq.~(\ref{eq:separ}) for an accurate representation of the 
spectral function. These eigenfunctions are in principle sp 
overlap functions (see discussion after Eq.~(\ref{eq:myspec}) 
below). They can be thought of as the 
natural orbits at a given energy. In fact, if the diagonalization 
is performed after integrating over the energy $E$ one would 
precisely obtain the natural orbits associated with the one-body 
density matrix and the eigenvalues $S_{lj}(i)$ would be the 
natural occupation numbers \cite{po95}.

\section{General formalism of DWIA}

For the scattering of an ultrarelativistic electron with initial
(final) momentum ${\bf p}_{\rm e} \
({\bf p}'_{\rm e})$, while a nucleon is ejected with final
momentum ${\bf p}'_{\rm N}$, the differential cross
section in the one-photon exchange approximation
reads~\cite{bgp82,bgprep}

\begin{equation}
{ {{\rm d}\sigma} \over {{\rm d}{\bf p}'_{\rm e}
{\rm d}{\bf p}'_{\rm N} } }= { e^4 \over {16 \pi^2}} {1 \over
Q^4 p_{\rm e} p'_{\rm e} } \
\lower7pt\hbox{$_{\lambda,\lambda'=0,\pm 1}$} \kern-28pt 
{\hbox{\raise2.5pt \hbox{$\sum$}}} 
\quad \  L_{\lambda,\lambda'}
W_{\lambda,\lambda'} , \label{eq:cross}
\end{equation}
where $Q^2 = {\bf q}^2 - \omega^2$ and ${\bf q} =
{\bf p}_{\rm e} - {\bf p}'_{\rm e}, \  \omega =
p^{}_{\rm e} - p'_{\rm e}$ are the momentum and energy
transferred to the target nucleus, respectively. The quantities
$L_{\lambda,\lambda'}, W_{\lambda,\lambda'}$ (usually referred to
as the lepton and hadron tensors, respectively) are expressed
in the basis of unit vectors

\begin{eqnarray}
e_0 &= &\left( 1, 0, 0, 0 \right) , \nonumber \\
e_{\pm 1} &= &\left( 0, \mp {\textstyle \sqrt{{1 \over 2}}}, -
{\textstyle \sqrt{{1 \over 2}}} {\rm i}, 0 \right) ,
\label{eq:basis}
\end{eqnarray}
which define the longitudinal (0) and transverse $(\pm 1)$
components of the nuclear response with respect to the
polarization of the exchanged virtual photon. The components
of the lepton tensor depend only on the electron kinematics,
while $W_{\lambda,\lambda'}$ depend on $q, \omega,
p'_{\rm N}, \cos \gamma = {\bf p}'_{\rm N} \cdot
{\bf q} / p'_{\rm N} q,$ and the angle $\alpha$ between the
$({\bf p}'_{\rm N},{\bf q})$ plane and the electron
scattering plane.

The hadron tensor is defined as~\cite{bgp82,bgprep,frumou}

\begin{equation}
W^{}_{\lambda,\lambda'} = \  
{\lower7pt\hbox{$_{\rm i}$}} \kern-7pt {\hbox{\raise7.5pt
\hbox{$\overline \sum$}}} \
\hbox {\hbox {$\sum$} \kern-15pt
{$\displaystyle \int_{\rm f}$\ } }
J^{}_{\lambda} ({\bf q}) J^*_{\lambda'} ({\bf q}) \  \delta
\left( E^{}_{\rm i} - E^{}_{\rm f} \right) ,
\label{eq:hadtens}
\end{equation}
i.e. it involves the average over initial states and the sum over
the final undetected states (compatible with energy-momentum
conservation) of bilinear products of the scattering amplitude
$J_{\lambda} ({\bf q})$.

This basic ingredient of the calculation is defined as

\begin{equation}
J^{}_{\lambda} ({\bf q}) = \int {\rm d} {\bf r} \
{\rm e}^{i {\bf q}
\cdot {\bf r}} \langle \Psi^{\rm A}_{\rm f}\vert
{\hat J}^{}_{\mu} \cdot e^{\mu}_{\lambda} \vert \Psi^{\rm A}_0 
\rangle , \label{eq:scattampl}
\end{equation}
where the matrix element of the nuclear charge-current density
operator ${\hat J}_{\mu}$ is taken between the initial,
$\vert \Psi^{\rm A}_0 \rangle$, and the final,
$\vert \Psi^{\rm A}_{\rm f} \rangle$, nuclear states. A natural 
choice for $\vert \Psi^{\rm A}_{\rm f} \rangle$ is suggested by 
the experimental conditions of the reaction selecting a final 
state, which behaves asymptotically as a knocked out nucleon with 
momentum $p'_{\rm N}$ and a residual nucleus in a well-defined 
state $\vert \Psi^{{\rm A} - 1}_n (E) \rangle$ with energy $E$ 
and quantum numbers $n$. By projecting this specific channel out 
of the entire Hilbert space, it is possible to rewrite 
Eq.~(\ref{eq:scattampl}) in a one-body representation 
(in momentum space and omitting spin degrees of freedom for 
simplicity) as~\cite{bccgp82}

\begin{equation}
J^{}_{\lambda} ({\bf q}) = \int {\rm d} {\bf p} \
\chi^{\left( -\right)\, *}_{p'_{\scriptscriptstyle{\rm N}} E n}
({\bf p} + {\bf q}) \  {\hat J}^{\rm eff}_{\mu}
({\bf p}, {\bf q}) \cdot e^{\mu}_{\lambda} \
\phi^{}_{E n} ({\bf p}) [S_n(E)]^{1\over 2} , 
\label{eq:scattampl1}
\end{equation}
provided that ${\hat J}_{\mu}$ is substituted by an appropriate
effective one-body charge-current density operator
${\hat J}^{\rm eff}_{\mu}$, which guarantees the
orthogonality between $\vert \Psi^{\rm A}_0 \rangle$ and
$\vert \Psi^{\rm A}_{\rm f} \rangle$ besides taking into account 
effects due to truncation of the Hilbert space. Actually, the 
orthogonality defect is negligible in the standard kinematics 
for $(e,e'p)$ reactions and in DWIA ${\hat J}^{\rm eff}_{\mu}$ is 
usually replaced by a simple one-body current 
operator~\cite{bccgp82,br91,bgprep}.

The functions

\begin{eqnarray}
[S_n (E)]^{1\over 2} \phi^{}_{E n} ({\bf p}) &= 
&\langle \Psi^{{\rm A} - 1}_n (E) \vert a ({\bf p}) 
\vert \Psi^{\rm A}_0 \rangle , \nonumber \\
\chi^{\left( - \right)}_{p'_{\scriptscriptstyle{\rm N}} E n}
({\bf p}) &= &\langle \Psi^{{\rm A} - 1}_n (E) \vert a ({\bf p}) 
\vert \Psi^{\rm A}_{\rm f} \rangle
\label{eq:specampl}
\end{eqnarray}
describe the overlap between the residual state 
$\vert \Psi^{{\rm A} - 1}_n (E) \rangle$ and the hole produced in 
$\vert \Psi^{\rm A}_0 \rangle$ and 
$\vert \Psi^{\rm A}_{\rm f} \rangle$, respectively, by removing a 
particle with momentum ${\bf p}$. Both $\phi^{}_{E n},
\chi^{\left( - \right)}_{p'_{\scriptscriptstyle{\rm N}} E n}$
are eigenfunctions of a Feshbach-like nonlocal energy-dependent 
Hamiltonian referred to the residual nucleus, belonging to the 
eigenvalues $E$ and $E+\omega$, respectively~\cite{bc81}.
The norm of $\phi^{}_{E n}$ is 1 and $S_n (E)$ is the 
spectroscopic factor associated with the removal process, i.e. it 
is the probability that the residual nucleus can indeed be 
conceived as a hole produced in the target nucleus. The 
dependence of 
$\chi^{\left( - \right)}_{p'_{\scriptscriptstyle{\rm N}} E n}$
on $p'_{\rm N}$ is hidden in the asymptotic state 
$\vert \Psi^{\rm A}_{\rm f} \rangle$ and the boundary conditions 
are those of an incoming wave.

Because of the complexity of the eigenvalue problem in the
continuum, the Feshbach hamiltonian is usually replaced by a
phenomenological local optical potential $V({\bf r})$ of
the Woods-Saxon form with complex central and spin-orbit
components. It simulates the mean-field interaction between
the residual nucleus and the emitted nucleon with energy-dependent
parameters determined through a best fit of elastic
nucleon-nucleus scattering data including cross section
and polarizations. Then,
$\chi^{\left( - \right)}_{p'_{\scriptscriptstyle{\rm N}} E n}
\sim \chi^{\left( - \right)}_{p'_{\scriptscriptstyle{\rm N}}}$
is expanded in partial waves and a Schr\"odinger equation
including $V({\bf r})$ is solved for each component up
to a maximum angular momentum satisfying a $p'_{\rm N}$-dependent
convergency criterion~\cite{bgprep}. The nonlocality of the 
original Feshbach hamiltonian is taken into account by 
multiplying the optical-model solution by the appropriate Perey 
factor~\cite{perey}.

After summing over the undetected final states with quantum numbers
$n$ of the residual nucleus, the hadron tensor 
$W_{\lambda,\lambda'}$ in momentum space becomes

\begin{eqnarray}
W^{}_{\lambda,\lambda'} &\sim &\sum_n \int {\rm d} {\bf p}
{\rm d} {\bf p}' \
\chi^{\left( - \right) \, *}_{p'_{\scriptscriptstyle{\rm N}}}
({\bf p}+{\bf q}) {\hat J}^{}_{\mu} ({\bf p}, {\bf q}) \cdot
e^{\mu}_{\lambda} \  \phi^{}_{E n} ({\bf p})
\phi^*_{E n} ({\bf p}') S_n (E) \nonumber \\
&  &\hbox{\hskip 2cm}{\hat J}^{\dagger}_{\nu} ({\bf p}',
{\bf q}) \cdot e^{\nu \, \dagger}_{\lambda'} \
\chi^{\left( - \right)}_{p'_{\scriptscriptstyle{\rm N}}}
({\bf p}'+{\bf q}) \nonumber \\
&\equiv &\int {\rm d} {\bf p} {\rm d} {\bf p}' \
\chi^{\left( - \right) \, *}_{p'_{\scriptscriptstyle{\rm N}}}
({\bf p}+{\bf q}) {\hat J}^{}_{\mu} ({\bf p}, {\bf q}) \cdot
e^{\mu}_{\lambda} \  S ({\bf p}, {\bf p}'; E) \nonumber \\
&  &\hbox{\hskip 2cm}{\hat J}^{\dagger}_{\nu} ({\bf p}',
{\bf q}) \cdot e^{\nu \, \dagger}_{\lambda'} \
\chi^{\left( - \right)}_{p'_{\scriptscriptstyle{\rm N}}}
({\bf p}'+{\bf q}) , \label{eq:hadtens1}
\end{eqnarray}
where
\begin{equation}
S ({\bf p}, {\bf p}'; E) = \sum_n \  S_n (E) \phi^*_{E n} 
({\bf p}') \phi^{}_{E n} ({\bf p}) \label{eq:myspec}
\end{equation}
is the hole spectral function defined in Eq.~(\ref{eq:spec}).
Notice that the spin and isospin indices have been omitted for 
simplicity and the summation over $n$ is over the different 
partial wave contributions which are present at a given energy 
$E$. This sum should not be confused with the separable 
representation (Eq.~(\ref{eq:separ})) of the partial wave 
contributions to the spectral function $S_{lj}(p,p',E)$ defined 
in Eq.~(\ref{eq:specl}). Each $lj$-contribution, coming from 
either quasi-hole states (if $E$ is the correct excitation energy) 
or from states which are usually unoccupied in the standard 
shell model, can be separately computed, so that the total hadron 
tensor will look like
\begin{equation}
W^{}_{\lambda,\lambda'} \equiv \sum_{lj} \  
W^{lj}_{\lambda,\lambda'} \quad . \label{eq:hadlj}
\end{equation}

Experimental data for the $(e,e'p)$ reaction are usually 
collected as ratios between the measured cross section and 
$K \sigma_{\rm eN}$, where $K$ is a suitable kinematical factor 
and $\sigma_{\rm eN}$ is the elementary (half off-shell) 
electron-nucleon cross section. In this way the information 
contained in the five-fold differential cross section is reduced 
to a two-fold function of the missing energy 
$E_{\rm m} = \omega - T_{p'_{\scriptscriptstyle{\rm N}}} - 
E_{\rm x}$ ($T_{p'_{\scriptscriptstyle{\rm N}}}$ is the kinetic 
energy of the emitted nucleon and $E_{\rm x}$ is the excitation 
energy of the residual nucleus) and of the missing momentum
${\bf p}_{\rm m} = {\bf p}'_{\rm N} - {\bf q}$~\cite{data}.
Therefore, in the following Section results will be presented
under the form of the socalled reduced cross section~\cite{bgprep}

\begin{equation}
n ({\bf p}_{\rm m}) \equiv
{ {\rm d\sigma} \over {\rm d{\bf p}'_{\rm e}
{\rm d}{\bf p}'_{\rm N}} }  {1 \over {K \sigma_{\rm eN}}}
\quad . \label{eq:redcross}
\end{equation}

\section{Results}

In this Section we will discuss results for the reduced cross 
section defined in Eq.~(\ref{eq:redcross}) for $(e,e'p)$ 
reactions on $^{16}$O leading both to discrete bound states of the 
residual nucleus $^{15}$N and to states in the continuum at higher missing 
energy. Distortion of electron and proton waves has been taken into account 
through the effective momentum approximation~\cite{GP} and through the 
optical potential derived from the Schwandt parametrization~\cite{Schw} 
(see Tab. III in Ref.~\cite{leus94}), respectively. All results presented 
here have been obtained using the 
CC1 prescription\cite{cc1} for the half off-shell elementary 
electron-proton scattering amplitude in analogy with what has been 
commonly done in the analysis of the experimental data. 
We also employed the nonrelativistic description for this 
amplitude\cite{devan} to be consistent with the nonrelativistic 
calculation of the five-fold differential cross section.
In parallel kinematics, where most of the experimental data are available,
this choice does not produce very different results with respect to the
former, and, therefore, will not be considered in the following.

\subsection{Quasihole states}
In Fig.~\ref{fig:fig1} the experimental results 
for the transition to the ground state of $^{15}$N are 
displayed as a function of the missing momentum 
${\bf p}_{\rm{m}}$. These data points have been collected at 
NIKHEF choosing the socalled parallel kinematics\cite{leus94}, 
where the direction of the momentum of the outgoing proton, 
${\bf p}_{\rm{N}}'$, has been fixed to be parallel to the 
momentum transfer ${\bf q}$. In order to minimize the effects of 
the energy dependence of the optical potential describing the 
FSI, the data points have been collected at a constant kinetic 
energy of 90 MeV in the center-of-mass system of the emitted 
proton and the residual nucleus. Consequently, since the momentum 
of the ejected particle is also fixed and
\begin{equation}
p_{\rm m} = \vert p_{\rm{N}}' \vert - \vert q \vert ,
\label{eq:missmo} 
\end{equation}
the missing momentum can 
be modified by collecting data at various momenta $q$ transferred
from the scattered electron. 

The experimental data points for this reduced cross section are
compared to the predictions of the calculations discussed above. 
The quasihole part of the spectral function for the 
$p{1 \over 2}$ partial wave represents the relevant piece of the 
nuclear structure calculation for the proton knockout reaction 
leading to the ground state of $^{15}$N. Using the quasihole part 
of the spectral function as discussed above (see 
Eq.~(\ref{eq:skoqh})) but adjusting the spectroscopic factor for 
the quasihole state contribution $Z_{0p{1 \over 2}}$ to fit the 
experimental data, we obtain the solid line of 
Fig.~\ref{fig:fig1}. Comparing this result with the experimental 
data one finds that the calculated spectral function reproduces 
the shape of the reduced cross section as a function of the 
missing momentum very well. The absolute value for the reduced 
cross section can only be reproduced by assuming a spectroscopic 
factor $Z_{0p{1 \over 2}} = 0.644$, a value considerably below 
the one of 0.89 calculated from Eq.~(\ref{eq:qhs})\cite{mu95}.
The phenomenological Woods-Saxon wave functions adjusted to fit 
the shape of the reduced cross section require spectroscopic 
factors ranging from 0.61 to 0.64 for the lowest 
$0p \textstyle{1 \over 2}$ state and from 0.50 to 0.59 for the 
$0p \textstyle{3 \over 2}$ state, respectively, 
depending upon the choice of the optical potential for the 
outgoing proton\cite{leus94}. The fact that the calculated 
spectroscopic factor is larger than the one adjusted to the 
experimental data may be explained by the observation that the 
calculations only reflects the depletion of the quasihole 
occupation due to short-range correlations. Further depletion and 
fragmentation should arise from long-range correlations due to 
collective excitations at low energies\cite{geu96,nili96}. Other 
explanations for this discrepancy could be the need for improving 
the description of spurious center-of-mass motion\cite{pwp,rad94} 
or a different treatment of FSI in terms of a relativistic model 
for the optical potential\cite{udias}.

In order to visualize the effects of FSI, Fig.~\ref{fig:fig1} 
also displays the results obtained for the quasihole contribution 
to the spectral function (with the same spectroscopic factor 
$Z_{0p{1 \over 2}} = 0.644$ as before, for sake of consistency) 
but ignoring the effects of the optical potential. In this 
socalled Plane-Wave Impulse Approximation (PWIA) the reduced cross 
section as a function of the missing momentum is identical to the 
spectral function at the missing energy of the considered 
$0p{1 \over 2}$ state, or, better, to the momentum distribution of 
the peak observed at this missing energy with the quantum numbers 
of the ground state of $^{15}$N. Therefore, the difference 
between the solid and the dashed line in Fig.~\ref{fig:fig1} 
corresponds to the difference between the reduced cross section 
defined in Eq.~(\ref{eq:redcross}) and the momentum distribution 
for the ground state of $^{15}$N. In other words, it illustrates
the effect of all the ingredients entering the present theoretical 
description of the $(e,e'p)$ reaction, which are not contained 
in the calculation of the spectral function. In particular,
the real part of the optical potential yields a 
reduction of the momentum of the outgoing proton $p_{\rm{N}}'$.
According to Eq.~(\ref{eq:missmo}), this implies in parallel
kinematics a redistribution of the strength towards smaller values
of the missing momentum and makes it possible to reproduce the observed
asymmetry of the data around $p_{\rm{m}} = 0$.
This feature cannot be obtained in PWIA (dashed line), where the results
are symmetric around $p_{\rm{m}} = 0$ due to the cylindrical symmetry of the
hadron tensor $W^{}_{\lambda,\lambda'}$
around the direction of ${\bf q}$ when FSI are switched off
(for a general review see Ref.~\cite{bgprep} and references therein).
The imaginary part of the optical potential describes 
the absorption of the proton flux due to coherent inelastic 
rescatterings, which produces the well known quenching with 
respect to the PWIA result.

As a second example for the reduced cross section in $(e,e'p)$
reactions on $^{16}$O leading to bound states of the residual 
nucleus, we present in Fig.~\ref{fig:fig2} the data for the 
$\textstyle{3 \over 2}^-$ state of $^{15}$N at an excitation 
energy of $-6.32$ MeV. 
Also in this case the experimental data are reproduced very well 
if we adjust the spectroscopic factor for the corresponding 
quasihole part in the spectral function to $Z_{0p{3 \over 2}} = 
0.537$. The discrepancy with the calculated spectroscopic factor 
(0.914) is even larger for this partial wave than it is for the 
$p{1 \over 2}$ state. A large part of this discrepancy can be 
attributed to the long-range correlations, which are not 
accounted for in the present study. Note, that in the experimental
data three $\textstyle{3 \over 2}^-$ states are observed in 
$^{15}$N at low 
excitation energies. Long-range correlations yield a splitting 
such that 86\% of the total strength going to these 
three states is contained in the experimental data displayed in 
Fig.~\ref{fig:fig2}. This splitting is not observed in the 
theoretical calculations. If one divides the adjusted 
spectroscopic factor $Z_{0p{3 \over 2}}$ by 0.86 to account for 
the splitting of the experimental strength, one obtains a value 
of 0.624 which is close to the total spectroscopic factor 
adjusted to describe the knockout of a proton from $p{1 \over 2}$ 
state.

Figure \ref{fig:fig2} also contains the results for the reduced 
cross section derived by substituting the overlap 
$[S_n (E)]^{1\over 2} \phi^{}_{E n}$ in Eq.~(\ref{eq:scattampl1}) 
with the variational wave function of Pieper et al.\cite{rad94}, 
who employed the Argonne potential for the NN 
interaction\cite{argon}. Also in this case the shape of the 
experimental data is globally reproduced with a slightly better 
agreement for small negative values of $p_{\rm m}$ but with a 
clear underestimation at larger $p_{\rm m}$. The overall quality 
of the fit is somewhat worse than for the Green's function approach 
and the required adjusted spectroscopic factor is 
$Z_{0p{3 \over 2}} = 0.459$, even below the value of 0.537 needed 
in the present calculation.

The analysis of the reduced cross section has been extended to 
higher missing momenta by experiments performed at the MAMI 
accelerator in Mainz\cite{blo95}, adopting different kinematical 
conditions than the parallel kinematics. Using the same 
spectroscopic factors for the $p{3 \over 2}$ and the 
$p{1 \over 2}$ partial waves, which were adjusted to the NIKHEF 
data above, the results of our calculations agree quite well 
also with these MAMI data, as displayed in Fig.~\ref{fig:fig2a}.
Although the calculation is somewhat below the data at high 
missing momentum, one should keep in mind that the corresponding 
difference in sp strength is only an extremely tiny fraction of 
the 10\% of the protons which are expected to be associated with 
high momenta due to short-range 
correlations~\cite{mudi94,mu95,po95}.

\subsection{The contribution of the continuum}
>From theoretical studies it is known that an enhancement of the
high-momentum components due to short-range NN correlations
does not show up in knockout experiments leading to states of low
excitation energy in the (A-1) nucleus, but should be seen at 
higher missing energies, which correspond to large excitation
energies in the residual nucleus. A careful analysis of 
such reactions leading to final states above the threshold for 
two-nucleon emission, however, is much more involved. For 
example, a description of the electromagnetic vertex beyond the 
impulse approximation is needed and two-body current operators 
must be adopted which are consistent with the contributions 
included in the spectral function. Moreover, the possible further 
fragmentation of the (A-1) residual system requires, for a 
realistic description of FSI, a coupled-channel formalism with 
many open channels. Calculations based on the optical potential 
are not satisfactory at such missing energies, because 
inelastic rescatterings and multi-step processes will add and 
remove strength from this particular channel. 

Nevertheless, it should be of interest to analyze the predictions 
of the present approach at such missing energies. First of all, 
because it represents the first realistic attempt of a complete 
calculation of the single-particle channel leading to the final 
proton emission, including intermediate states above the Fermi 
level up to $l=4$; therefore, it represents a realistic estimate 
of the relative size of this specific channel. Secondly, because 
information on the shape of the reduced cross section as a
function of the missing momentum or on the relative contribution of
various partial waves could yield reliable results even at these
missing energies. Due to the problems mentioned above, no 
reliable description of the absolute value of the reduced cross
section can be reached in this framework.

In order to demonstrate the energy dependence of the spectral 
function and its effect on the cross section, we have calculated 
the reduced cross section for the excitation of 
$\textstyle{3 \over 2}^-$ states 
at $E_{\rm m} = -63$ MeV. For these studies we considered 
the socalled perpendicular kinematics, where the energy of the 
emitted proton is kept fixed at 90 MeV as well as the momentum 
transfer at $q \sim 420$ MeV/c (equal to the outgoing proton 
momentum). The same optical potential as in Figs.~\ref{fig:fig1}, 
\ref{fig:fig2} can be adopted to describe FSI and the 
missing momentum distribution is obtained by varying the angle 
between ${\bf p}_{\rm N}'$ and ${\bf q}$. For a spectral function 
normalized to unity (as the absolute result for the cross section 
is not reliable), the reduced cross section is represented by the 
solid line in Fig.~\ref{fig:fig3}. If, however, we replace the 
spectral function derived from the continuum contribution in 
Eq.~(\ref{eq:renor}) by the one derived for the 
$\textstyle{3 \over 2}^-$ 
quasihole state at its proper missing energy (but now in the same 
kind of perpendicular kinematics and normalized to 1)  
we obtain the dashed line. A comparison of these two calculations 
demonstrates the enhancement of the high-momentum components in 
the spectral function leading to final states at large excitation 
energies. Note that the cross section derived from the appropriate 
spectral function is about two orders of magnitude larger at 
$p_{\rm m} \sim 500$ MeV/c than the one derived from the spectral 
function at the quasihole energy.

The discussion so far is of course somewhat academic since it will
be difficult to perform a decomposition of the continuum 
contribution to the reduced cross section in terms of the quantum 
numbers for angular momentum and parity of the state for
the residual system. Therefore we display in
Figs.~\ref{fig:fig4} and \ref{fig:fig5} the contributions to the 
total reduced cross section of the various partial waves 
associated to states above the Fermi level and usually unoccupied 
in the standard shell model. From Fig.~\ref{fig:fig4} we can 
furthermore see that the relative importance of the various 
partial waves changes with the missing momentum, emphasizing the 
contribution of higher angular momenta at increasing $p_{\rm m}$. 
This feature can be observed even better in Fig.~\ref{fig:fig5}, 
where the percentage of each relative contribution to the total 
reduced cross section is displayed as a function of the missing 
momentum. For each orbital angular momentum we obtain a 
``window'' in $p_{\rm m}$ where its contribution shows a maximum 
as compared to other partial waves. 

\section{Conclusions}

In the present paper the consequences of the presence of 
high-momentum components in the ${}^{16}$O ground state have been 
explored in the calculation of the $(e,e'p)$ cross section within 
the formalism for the DWIA developed in 
Refs.\cite{bc81,bccgp82,bgp82,br91,bgprep,libro}. The spectral 
functions have been calculated for the ${}^{16}$O system itself, 
by employing the techniques developed and discussed 
in\cite{boro,mudi94,mu95,po95}. At low missing energies, the 
description of the missing momentum dependence of the 
$p \textstyle{1 \over 2}$ and $p \textstyle{3 \over 2}$ 
quasihole states compares favorably with the experimental data 
obtained at NIKHEF\cite{leus94} and at the MAMI facility in 
Mainz\cite{blo95}. The difference between theory and experiment 
at high missing momenta can at most account for a very tiny 
fraction of the sp strength which is predicted to be present at 
these momenta\cite{mudi94,mu95,po95}. A comparison with the PWIA 
result clarifies the influence of FSI in parallel kinematics.
We also compare our results for the $p \textstyle{3 \over 2}$ 
quasihole state with the results obtained in Ref.~\cite{rad94} 
for the Argonne NN interaction. While the shape of the cross 
sections is nicely described by our results, the associated 
spectroscopic factors are overestimated substantially. Although a 
large fraction of this discrepancy can be ascribed to the 
influence of long-range correlations\cite{geu96,nili96}, which 
are outside the scope of the present work, a discrepancy
may still remain although it has been suggested 
that a correct treatment of the center-of-mass 
motion\cite{rad94} may fill this gap.

As discussed previously for nuclear matter (see $e.g.$ 
\cite{cio91}) and emphasized in \cite{mudi94,mu95,po95} for 
finite nuclei, the admixture of high-momentum components in the 
nuclear ground state can only be explored by considering high 
missing energies in the $(e,e'p)$ reaction. Although other 
processes may contribute to the cross section at these energies, 
we have demonstrated in this paper that the expected
emergence of high missing momentum components in the cross section
is indeed obtained and yields substantially larger cross sections
than the corresponding outcome for the quasihole states.
As a result, we conclude that the presence of high-momentum 
components leads to a detectable cross section at high missing 
energy. In addition, we observe that it is important to include 
orbital angular momenta at least up to $l = 4$ in the spectral 
function in order to account for all the high missing momentum 
components up to about 600 MeV/c. A clear window for the dominant 
contribution of each $l$-value as a function of missing 
momentum is also established. This feature may help to analyze 
experimental data at these high missing energies.

\vspace{1cm}
\centerline{\bf ACKNOWLEDGEMENTS}

This research project has been supported in part by Grant No. 
DGICYT, PB92/0761 (Spain), EC Contract No. CHRX-CT93-0323, the
``Graduiertenkolleg Struktur und Wechselwirkung von Hadronen und 
Kernen'' under DFG Mu705/3 (Germany), and the U.S. NSF under 
Grant No. PHY-9602127.


\begin{figure}
\caption{Graphical representation of the Hartree-Fock (a),
the two-particle-one-hole (2p1h,b) and the two-hole-one-particle
contribution (2h1p,c) to the self-energy of the nucleon.}
\label{fig:diag}
\end{figure}
\begin{figure}
\caption{Reduced cross section for the 
$^{16}$O$(e,e'p)^{15}$N$_{\rm gs}$ reaction in parallel 
kinematics. Results with 
(solid line) and without (dashed line) inclusion of the FSI are 
compared to the experimental data\protect\cite{leus94}. A 
spectroscopic factor of 0.644 has been employed in displaying the
results for the calculations involving the spectral function.}
\label{fig:fig1}
\end{figure}
\begin{figure}
\caption{Reduced cross section for the $^{16}$O$(e,e'p)$ reaction 
in parallel kinematics leading to the $\textstyle{3 \over 2}^-$ 
state at $-6.32$ MeV of the residual nucleus $^{15}$N. Results of  
the present Green's function approach (solid line) are compared 
to those obtained in the variational calculation of 
\protect\cite{rad94} (dashed line) and the experimental 
data\protect\cite{leus94}. A spectroscopic 
factor of 0.537 was required for the Green's function approach, 
while $Z_{p \scriptstyle{3 \over 2}} = 0.459$ has been used to 
adjust the results of the variational calculation.}
\label{fig:fig2}
\end{figure}
\begin{figure}
\caption{Reduced cross section for the $^{16}$O$(e,e'p)$ reaction 
leading to the ground and the 
$\textstyle {3 \over 2}^-$ states of $^{15}$N in the 
kinematical conditions considered in the experiment of 
\protect\cite{blo95}. The calculations were performed using the 
same spectral functions as discussed for Figs.~2 and 3.}
\label{fig:fig2a}
\end{figure}
\begin{figure}
\caption{Reduced cross section for the $^{16}$O$(e,e'p)$ reaction 
in perpendicular kinematics for the excitation of 
$\textstyle {3 \over 2}^-$ states at $E_{\rm m} = -63$ MeV 
(solid line) and $-6.32$ MeV (dashed line).}
\label{fig:fig3}
\end{figure}
\begin{figure}
\caption{Contributions of various partial waves to the reduced 
cross section for the $^{16}$O$(e,e'p)$ reaction in the same 
conditions as for the solid line in Fig. 5.}
\label{fig:fig4}
\end{figure}
\begin{figure}
\caption{Relative importance of various partial waves to the 
reduced cross section for the $^{16}$O$(e,e'p)$ reaction 
in the same conditions as in Fig. 6.}
\label{fig:fig5}
\end{figure}
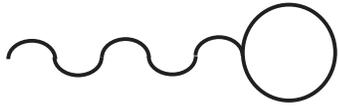
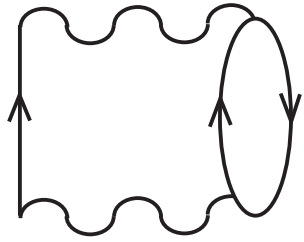
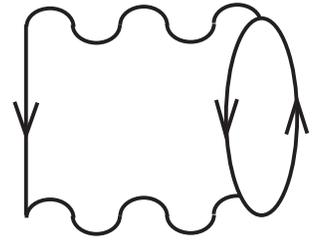
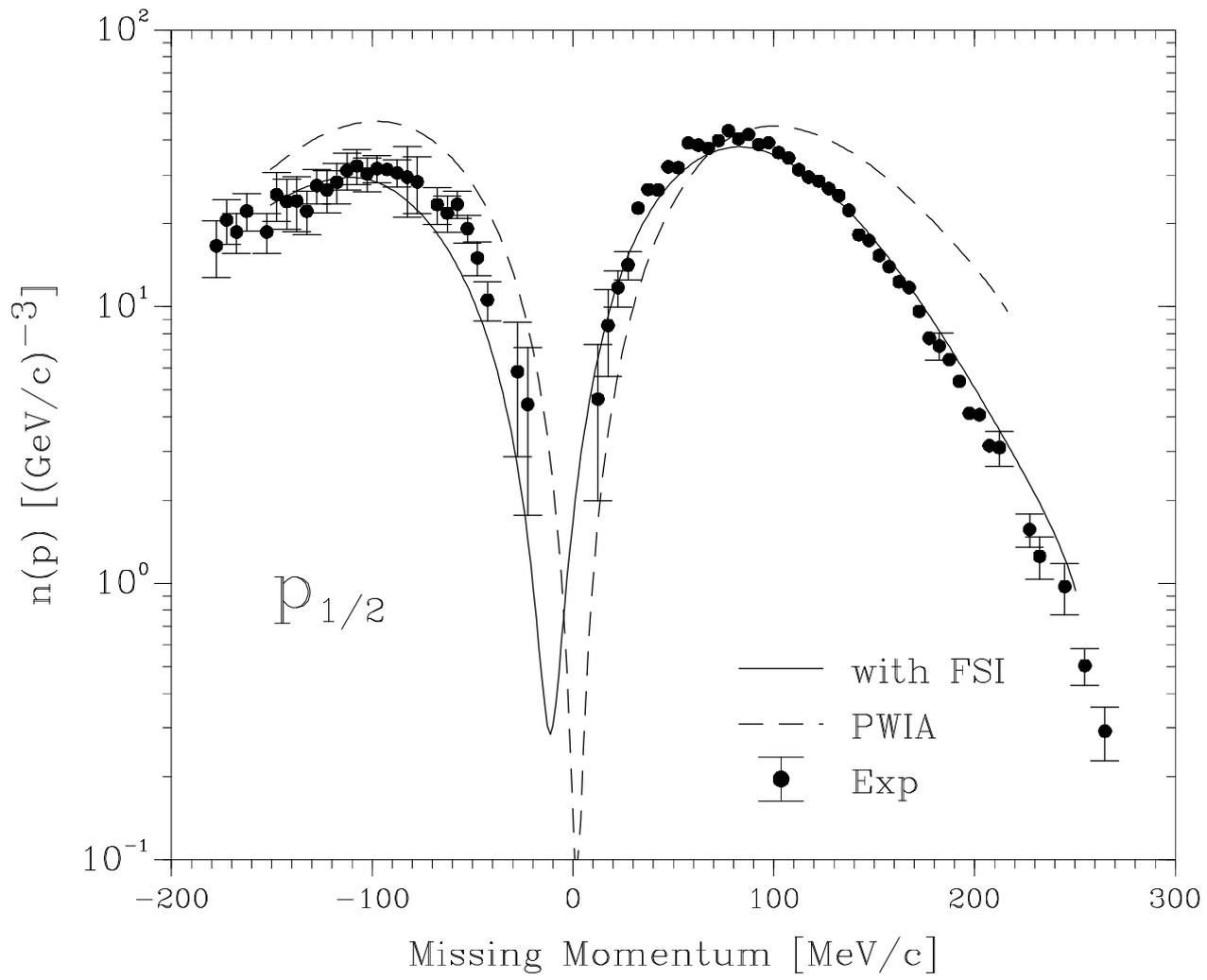
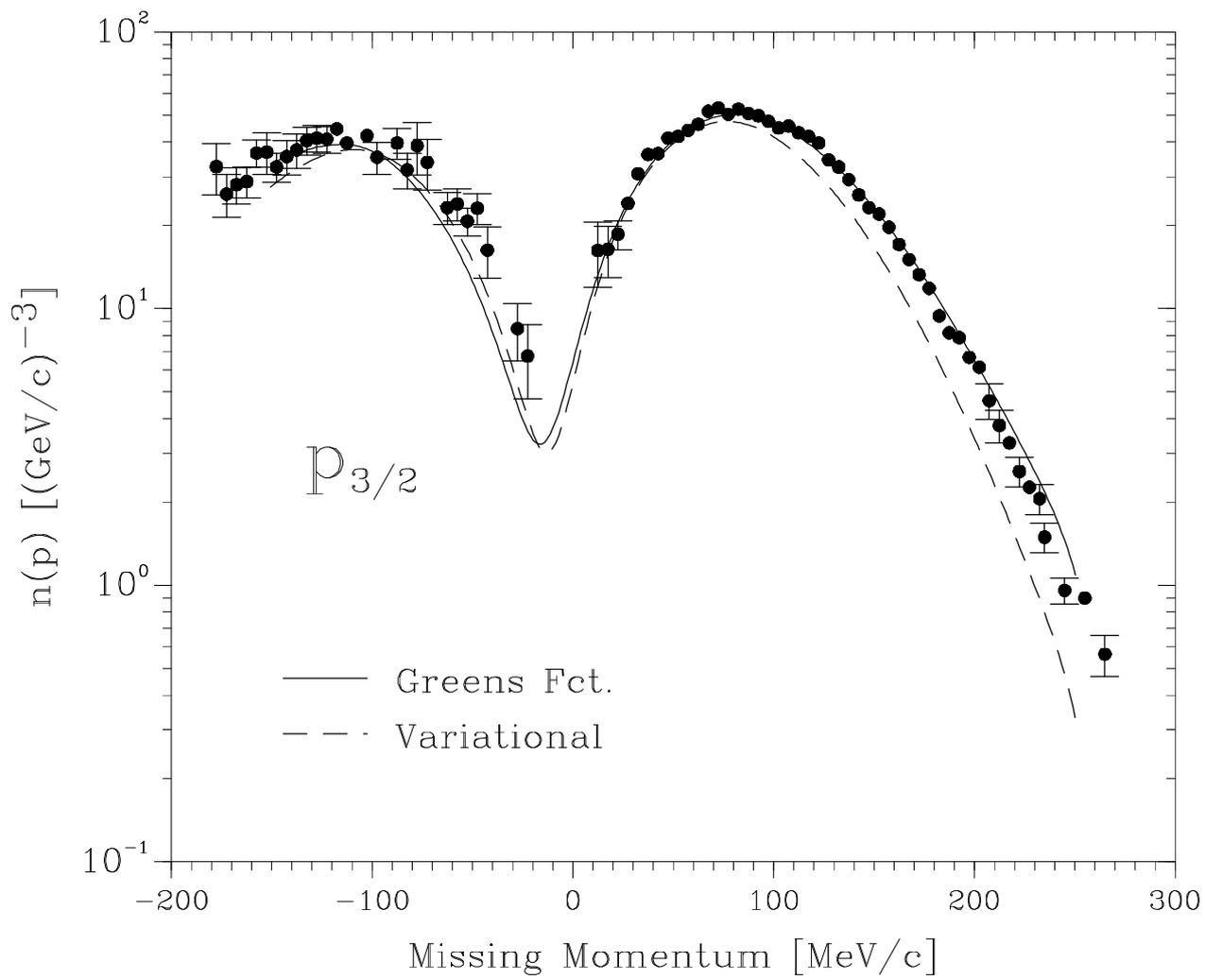
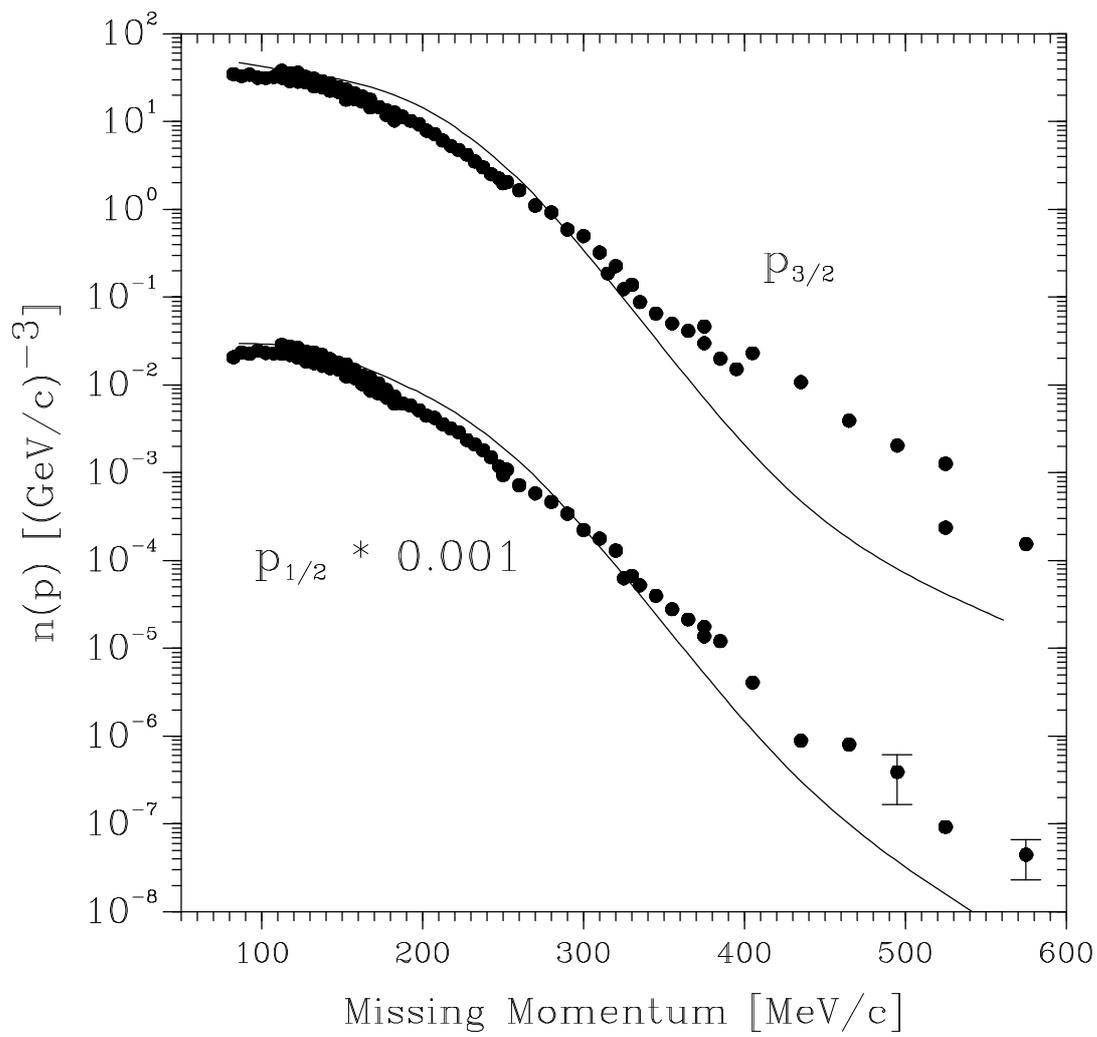
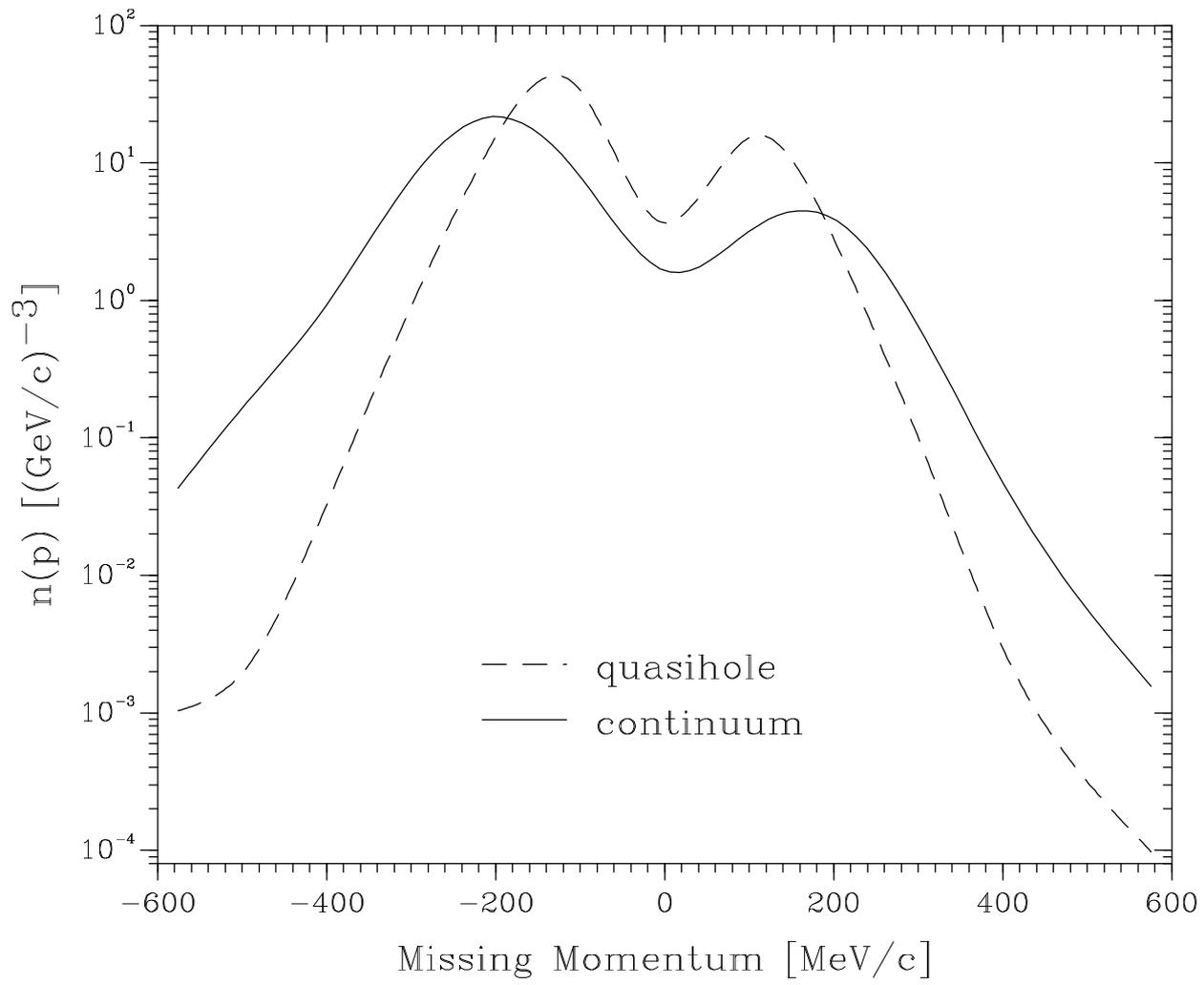
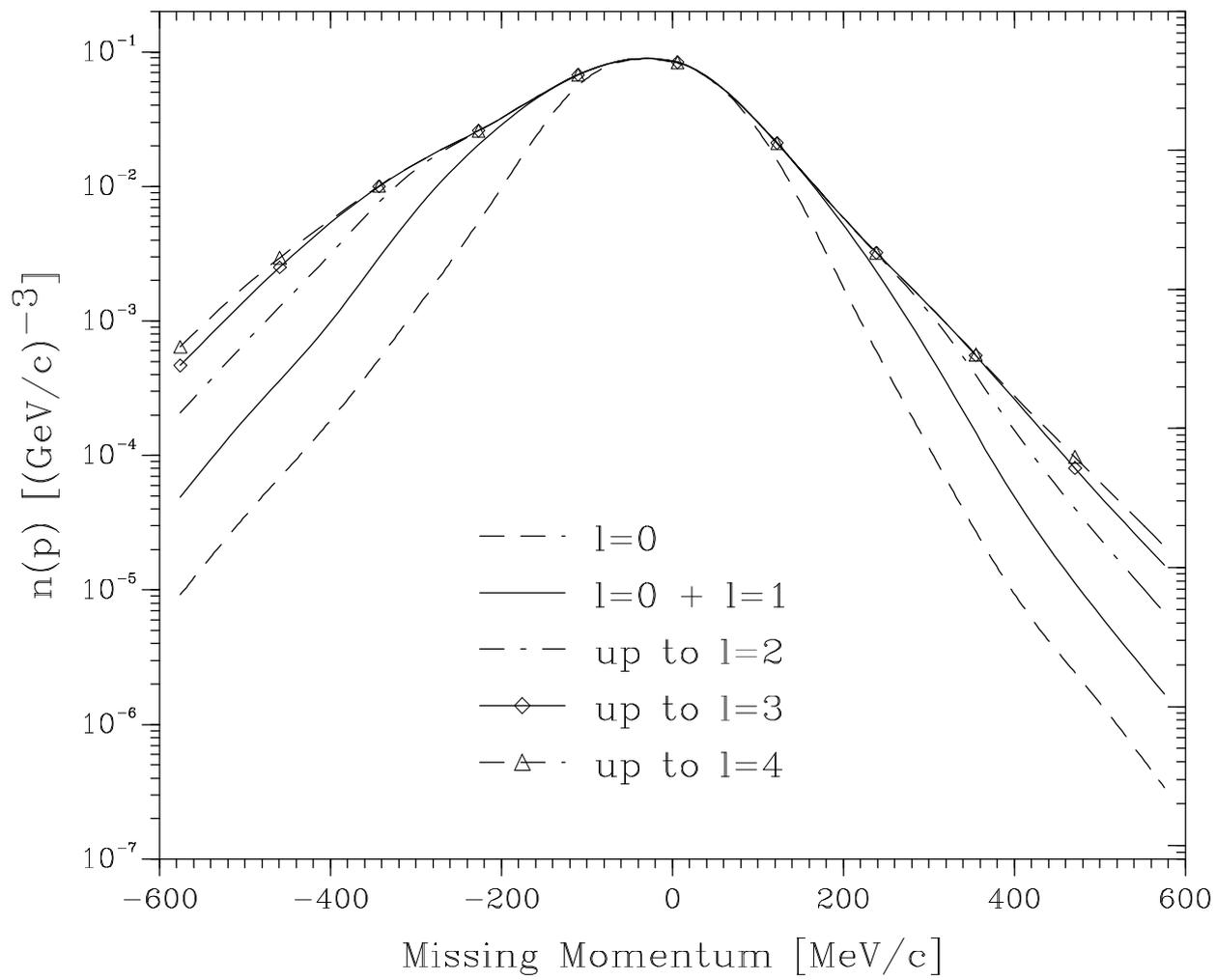
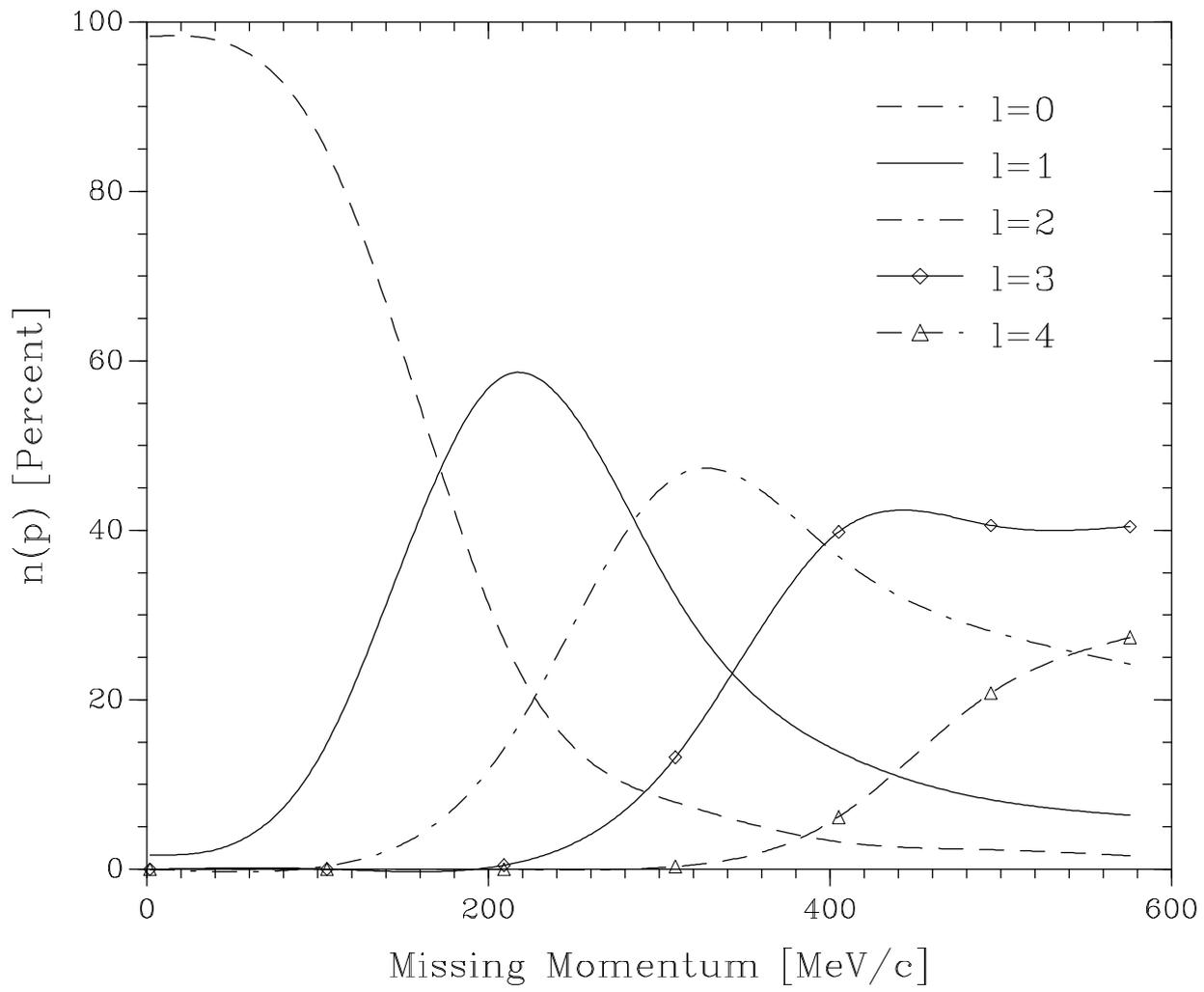
\end{document}